\documentclass[
  twoside,
  twocolumn,
  prb,
  preprintnumbers,
  floatfix,
  showpacs, superscriptaddress,
  ]{revtex4-1}
\usepackage{amsmath}
\usepackage{amssymb}
\usepackage{color}
\usepackage{graphicx}
\usepackage{tabularx}
\usepackage{dcolumn} 
\usepackage{bm}      
\usepackage{array}   
\usepackage{amsfonts}
\usepackage{times}

\newcolumntype{.}[1]{D{.}{.}{#1}}

\begin{document}

\title{Field-Dependent Qubit Flux Noise Simulated from Materials-Specific Disordered Exchange Interactions Between Paramagnetic Adsorbates}

\author{Keith G. Ray}
\email{ray30@llnl.gov}
\author{Yaniv Rosen}
\author{Jonathan L Dubois}  
\author{Vincenzo Lordi}  
\email{lordi2@llnl.gov}
\affiliation{Lawrence Livermore National Laboratory, Livermore, CA 94550, USA}

\begin{abstract}

Superconducting quantum devices, from qubits and magnetometers to dark matter detectors, are influenced by magnetic flux noise originating from paramagnetic surface defects and impurities.  These spin systems can feature complex dynamics, including a Berezinskii-Kosterlitz-Thouless transition, that depend on the lattice, interactions, external fields, and disorder. However, the disorder included in typical models is not materials-specific, diminishing the ability to accurately capture measured flux noise phenomena. We present a first principles-based simulation of a spin lattice consisting of paramagnetic O$_2$ molecules on an Al$_2$O$_3$ surface, a likely flux noise source in superconducting qubits, to elucidate opportunities to mitigate flux noise. We simulate an ensemble of surface adsorbates with disordered orientations and calculate the orientation-dependent exchange couplings using density functional theory. Thus, our spin simulation has no free parameters or assumed functional form of the disorder, and captures correlation in the defect landscape that would appear in real systems. We calculate a range of exchange interactions between electron pairs, with the smallest values, 0.016 meV and -0.023 meV, being in the range required to act as a two-level system and couple to GHz resonators. We calculate the flux noise frequency, temperature, and applied external magnetic field dependence, as well as the susceptibility-flux noise cross-correlation. Calculated trends agree with experiment, demonstrating that a surface harboring paramagnetic adsorbates arranged with materials-specific disorder and interactions captures the various properties of magnetic flux noise observed in superconducting circuits. In addition, we find that an external electric field can tune the spin-spin interaction strength and reduce magnetic flux noise.

\end{abstract}

\maketitle


Magnetic flux noise makes up one of the primary sources of decoherence in superconducting quantum information devices.\cite{yoshihara2006decoherence,clarke2008superconducting,anton2013magnetic} The magnetic flux noise couples to superconducting loops, introducing dephasing in flux and phase qubits, as well as flux-tuned qubits.\cite{clarke2008superconducting,paladino20141,hutchings2017tunable} Many properties of this noise source have been determined experimentally. The magnetic flux noise power has been measured to have a roughly 1/$f$ frequency dependence over a range of frequencies from sub-Hz to GHz\cite{koch1983flicker,yoshihara2006decoherence,Quintana2017ClassicalQuantumCrossoverFluxNoise} and has a significant cross-correlation with susceptibility that decreases with temperature.\cite{sendelbach2009complex} Furthermore, the classical-quantum crossover of the flux noise was observed and the antisymmetric component of the flux noise was found to have a 1/T dependence, suggesting a paramagnetic environment.\cite{Quintana2017ClassicalQuantumCrossoverFluxNoise} A cusp in the flux vs.\ temperature for an 870~pH Nb/AlO$_\text{x}$/Nb SQUID, reminiscent of cusps in the first-order susceptibility of a spin glass at the freezing temperature, was measured at 55~mK by Sendelbach et al.\cite{sendelbach2008magnetism} In addition, Lanting et al. argued with spin diffusion simulations and measurements of rf SQUID flux qubits that a spin system near the spin-glass phase transition can be used to interpret the frequency and temperature dependence of flux noise measurements.\cite{Lanting2014SpinDiffusionFluxNoise}  However, the presence of a spin glass alone, with zero net magnetization, is inconsistent with the measured non-zero flux noise--susceptibility cross-correlation.

Adsorbed paramagnetic species have been proposed as the origin of the flux noise. For example, X-ray magnetic circular dichroism (XMCD) measurements have recorded a spin dependent absorption consistent with adsorbed oxygen.\cite{kumar2016origin} Other studies have observed signatures consistent with atomic hydrogen.\cite{de2017direct} Theoretical models have investigated paramagnetic adsorbates and defects,\cite{PhysRevLett.112.017001,Lordi2017} ferromagnetic clusters,\cite{de20191} O$_2$ on sapphire with a single exchange coupling and random distances,\cite{wang2015candidate} H on sapphire,\cite{wang2018hydrogen} spins coupled by RKKY interactions,\cite{faoro2008microscopic} metal induced gap states,\cite{choi2009localization} trapped electron states,\cite{Koch2007ModelFluxNoise}  and random couplings similar to a spin glass.\cite{atalaya2014flux} However, these models have either not included disorder or have done so only at the level of an assumed functional form describing the spread of exchange couplings. Real materials have disorder on a granular scale, which is set by the atomic spacing, that exhibits spatial correlations. These details are important to capturing the properties of experimentally measured magnetic flux noise, as well as developing rational mitigation schemes.

\begin{figure*}[!t]
\centering
\includegraphics[width=.98 \textwidth]{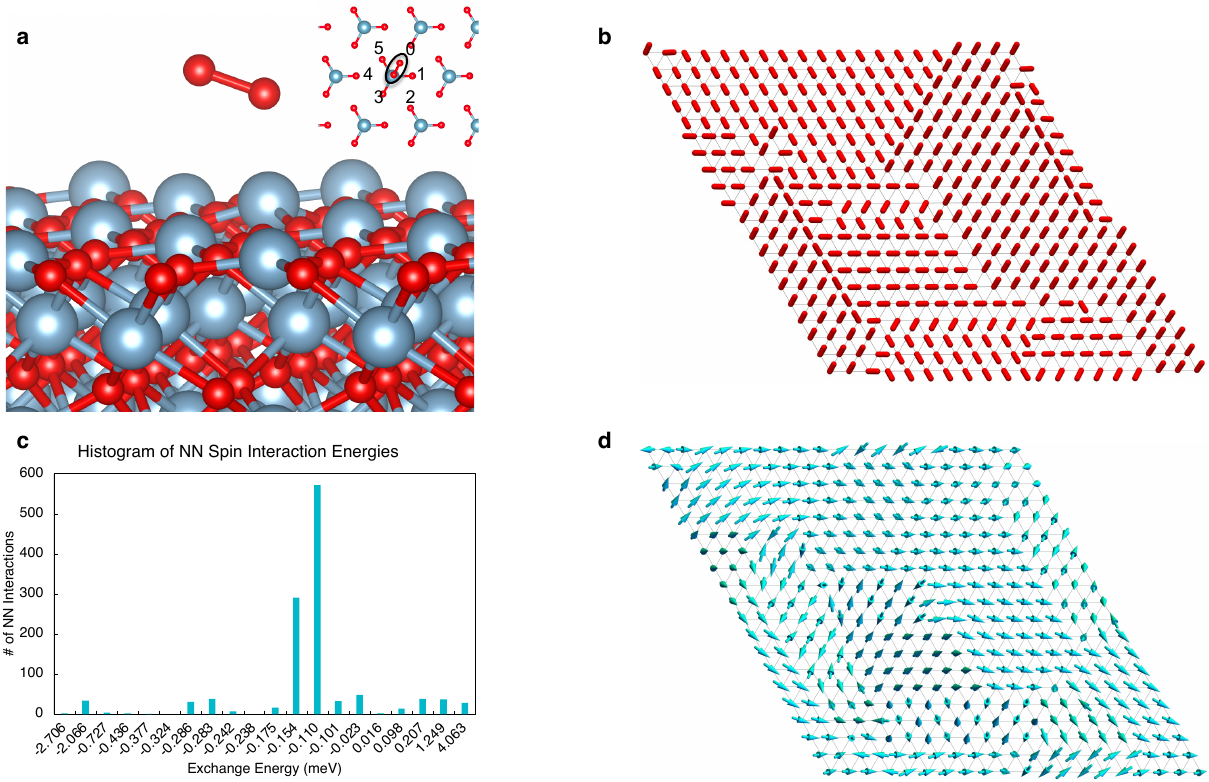}
\caption{(a) Oxygen molecule on the (0001) Al$_2$O$_3$ surface, inset: top-down view showing the six possible O$_2$ binding orientations. (b) Monte Carlo generated arrangement of an O$_2$ molecule monolayer at 0.01 K, where the O$_2$ dimers are represented by red cylinders with exaggerated length compared to the lattice spacing. (c) Distribution of O$_2$-O$_2$ spin exchange couplings from the arrangement depicted in (b), showing mostly weakly negative (ferromagnetic) interactions. (d) Monte Carlo generated lattice of O$_2$ spins, represented by light blue arrows, at 0.01 K on the O$_2$ molecular surface configuration shown in (b). 
}
\label{SpinLattice}
\end{figure*}

Surface magnetic flux noise sources are often governed by Berezinskii-Kosterlitz-Thouless (BKT) physics,\cite{berezinsky1970destruction,kosterlitz1973ordering} that is, they are systems that approximate the XY spin model. Therefore, understanding the BKT phase transition in these systems is important to understanding the flux noise. These spin systems are characterized by a lattice of interacting vectors, or spins, that each possess a circular symmetry, undergo a topological phase transformation between bound and unbound vortex-antivortex pairs, where below the transition temperature, $T_{\text{BKT}}$, the spin spatial correlation function decays algebraically.\cite{berezinsky1970destruction,kosterlitz1973ordering} This behavior is realized in a diverse set of intriguing and technologically-relevant materials, including the metal-insulator transition in disordered graphene,\cite{zhang2009localization} unconventional superconductivity in graphene superlattices,\cite{cao2018unconventional} trapped atomic gases,\cite{hadzibabic2006berezinskii} helium superfluids,\cite{minnhagen1987two} superconductor thin films,\cite{reyren2007superconducting,baturina2012superconducting,hebard1980evidence,kogan2007interaction} Jospehson junction arrays,\cite{martinoli2000two} polariton condensates,\cite{caputo2018topological} arrays of superconducting qubits,\cite{king2018observation} and two-dimensional magnets.\cite{sato2003unconventional}  The value of $T_{\text{BKT}}$, and also the degree of competition with other magnetic phases, is largely determined by the vector lattice geometry and sparseness, external magnetic field, and spin-spin interaction strength, as well as disorder in these parameters.\cite{kapikranian20072d, wysin2005extinction,maccari2017broadening,lee1998phase} In particular, disorder and frustration have been shown to affect $T_{\text{BKT}}$ and the vortex-core energy, broaden the superconducting transition in 2D materials, and induce a disorder-driven BKT-type metal-insulator transition.\cite{wysin2005extinction,xie1998kosterlitz,teitel1983phase, sato2003unconventional,maccari2017broadening,nomura2009field,mondal2011role,jose1981effects} In simulations of these systems, disorder is typically included by choosing lattice couplings, nearest neighbor separations, or occupancies according to a particular probability distribution. However, disorder considered in this manner does not capture the correlations between defects apparent in realistic systems. To do this, one must simulate an ensemble of materials configurations that the spin lattice lives upon.

The tuning of BKT transitions via external fields has  been explored. Magnetic fields have been used to alter the transition in Josephson junction arrays, from vortex-antivortex unbinding to a field-tuned vortex delocalization transition,\cite{fisher1990quantum,chen1995scaling} as well as superconductor-insulator transitions in amorphous thin films.\cite{hebard1990magnetic} An electrostatic field applied to a LaAlO$_3$/SrTiO$_3$ interface can tune the superconductivity in this system with properties consistent with a BKT transition under certain conditions.\cite{schneider2009electrostatically}  The tunability of the BKT transition could lead to new technologies or improvement of existing devices that involve BKT physics and work is being done on devices in general that have magnetic phases that can be tuned electronically.\cite{huang2018electrical,jiang2018controlling} Tuning strategies for the BKT transition might also allow for the tuning of the magnetic flux noise in paramagnetic spin systems.

In our study, we determine the magnetic fluctuations of an interacting system of paramagnetic spins residing on O$_2$ molecules adsorbed on an Al-terminated sapphire (0001) surface, as depicted in Fig.~\ref{SpinLattice}(a). Our modeling scheme incorporates disorder in the O$_2$ molecule orientations by employing a Monte Carlo simulation with the following Hamiltonian:
  \begin{equation}
  H_{\text{O}_2 \text{lattice}} = \sum_{i}  \bigg(E_i^b(O_i) + \frac{1}{2} \sum_{j=NN} F(O_i,O_j,\textbf{r}_i - \textbf{r}_j) \bigg),
  \end{equation}
where nearest neighbor (NN) O$_2$ interactions, $F(O_i,O_j,\textbf{r}_i - \textbf{r}_j)$, and the O$_2$ binding energies, $E_i^b(O_i)$, are both dependent on O$_2$ orientation, given by $O_{i}$ and $O_{j}$. We determined these explicit dependencies using van der Waals-corrected density functional theory (DFT) calculations.\cite{PhysRevB.89.035412} The energies, as well as details of the Monte Carlo simulations and DFT calculations are presented in the Supplemental Material. We consider 100\%, 75\%, and 50\% O$_2$ coverage and a typical disordered arrangement of O$_2$ orientations at 0.01~K is presented in Fig.~\ref{SpinLattice}(b) for a 20$\times$20 periodic lattice. The results we present in this work are for that size system. In the inset Fig.~\ref{SpinLattice}(a) the six stable O$_2$ orientations are given, which, with respect to the sapphire surface, can be divided into two sets of three-fold degenerate orientations separated by a small (16 meV) energy splitting. The quenched disorder in the adsorbed O$_2$ layer is frozen in below roughly 1~K, and so we employ 0.01~K for the O$_2$ ensembles to represent temperatures of 0 to 1~K without loss of generality.

The disorder in the oxygen molecule orientations causes there to be disorder in the magnetic exchange couplings between the paramagnetic oxygen spins due to different overlaps of the electron wavefunctions. We determine these couplings with DFT calculations and tabulate them in the Supplemental Material. For a 20$\times$20 periodic lattice of O$_2$ sites, there will be 400 O$_2$ sites, each with six nearest neighbors under full coverage conditions, for 1200 exchange interactions in the model. These 1200 exchange interactions will each have one of the 20  possible exchange interaction values arising from the number of possible symmetry-distinct relative nearest neighbor O$_2$ orientations on (0001) Al$_2$O$_3$.
 To illustrate the range of exchange couplings typically present, the histogram of the exchange couplings for the O$_2$ arrangement shown in Fig.~\ref{SpinLattice}(b) is shown in Fig.~\ref{SpinLattice}(c). Both ferromagnetic (FM) and antiferromagnetic (AFM) exchange energies, defined by negative and positive coupling parameters, respectively, are present with varying strengths from $-2.7$~meV to $+4.1$~meV, with $+0.016$~meV being the smallest exchange interaction in absolute magnitude and $-0.110$ meV and $-0.154$ meV being the most common. 
 We note that the $+0.016$~meV and $-0.023$~meV   exchange energies are small enough to couple to GHz resonators as a two-level system, potentially adding another loss channel. This TLS consists of a spin flip and corresponding electric dipole moment change, as described further on. Most of the exchange interactions realized with this O$_2$ arrangement, as well as with other O$_2$ arrangements in this study, are ferromagnetic, leading us to the picture of ferromagnetic clusters (domains) that are often separated by antiferromagnetic couplings (domain boundaries) for the full coverage case, consistent with other theoretical works in the literature.\cite{de20191,atalaya2014flux,PhysRevLett.104.247204,wang2015candidate}  For lower coverages of O$_2$,  ferromagnetic clusters persist, but are separated by additional unoccupied O$_2$ binding sites.

With an ensemble of simulated O$_2$ configurations defining the exchange interactions between their paramagnetic spins, we can now model the spin system that exists on the O$_2$ surface layers using the following Hamiltonian:
    \begin{multline}
  H_{\text{spin}} = \sum_{i}  \bigg(E_{\text{anisotropy}}(O_i,\textbf{s}_i) - (-g_{\text{s}}\mu_{\text{B}}\textbf{s}_i/\hbar) \cdot \textbf{H}_{\text{ext}} \\+ \frac{1}{2} \sum_{j=NN} J(O_i,O_j,\textbf{r}_i - \textbf{r}_j)\frac{\textbf{s}_i\cdot\textbf{s}_j}{2} \bigg),
  \label{Hspin}
  \end{multline}
where $i$ sums over all O$_2$ molecules, $j$ sums over nearest neighbors (NN), $E_{\text{anisotropy}}(O_i,\textbf{s}_i)$ defines the oxygen orientation-dependent anisotropy in the spin direction, $J(O_i,O_j,\textbf{r}_i - \textbf{r}_j)$ is the exchange coupling that depends on both oxygen orientation and direction of separation on the sapphire surface, and $\textbf{s}_i$ is the spin on the $i$th oxygen molecule from its two unpaired electrons. The spin anisotropy, given by $E_{\text{anisotropy}}(O_i,\textbf{s}_i) = 0.037~\text{meV} \times |\textbf{O}_i \cdot \textbf{s}_i|$, is such that the oxygen molecule unpaired spins are energetically favored to point in the plane perpendicular to the O$_2$ bond axis, $\textbf{O}_i$ for molecule $i$.  This term is responsible for this system approximating an XY model, as opposed to a Heisenberg model without spin anisotropy. A typical arrangement of surface spins generated with a Monte Carlo simulation, combining  Metropolis and Wolff Cluster update steps, is shown in Fig.~\ref{SpinLattice}(d), where several ferromagnetic domains, indicated by collinear spins, are observed. This arrangement of electron spins lives on the oxygen monolayer depicted in Fig.~\ref{SpinLattice}(b), and so we observe that domain walls in the spin lattice correspond to changes in O$_2$ orientation in the adsorbed monolayer.

\begin{figure}[h!]
\centering
\includegraphics[width=1.0 \columnwidth]{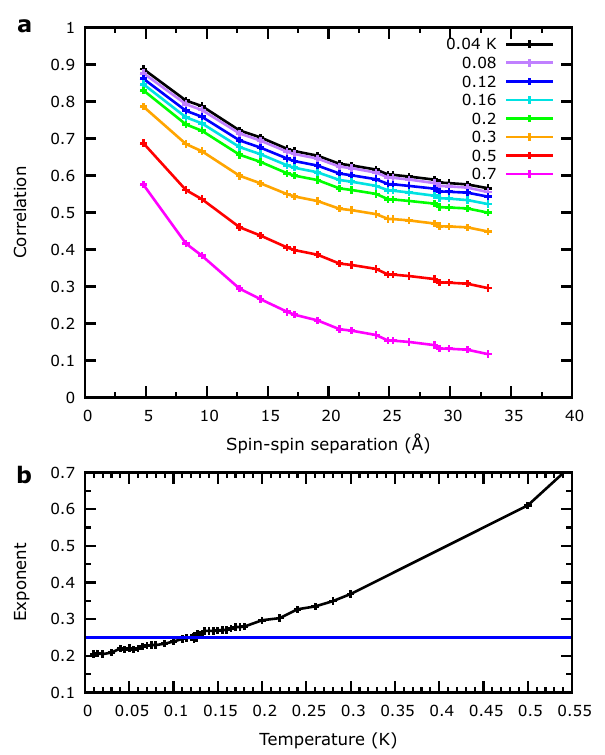}
\caption{(a) Spin-spin correlation functions for several temperatures show the algebraic decay characteristic of BKT phases below the transition temperature. (b) By fitting the correlation functions to the expected algebraic scaling, $(|r - r'|/l)^{-\eta}$, as a function of temperature, we can extract the transition temperature, which corresponds to the case with $\eta = 1/4$; the data here give a transition temperature of 121~mK.
}
\label{magT}
\end{figure}

Utilizing these methods, we are able to simulate the spin arrangements at varying temperature and under different applied magnetic fields.  As another indicator for the energy scales relevant to this system, the  BKT transition temperature can be extracted from the spin-spin correlation distance scaling, computed as a function of temperature, as shown in Fig.~\ref{magT}.  Below the transition temperature, the correlation function scales as $(|r - r'|/l)^{-\eta}$, where $r$ and $r'$ are the spin positions, $l$ is the lattice spacing (4.8~\AA~for the (0001) sapphire surface), and $\eta = {kT/2\pi |J|}$ is a scaling parameter, with $k$ being the Boltzmann constant, $T$  the temperature, and $J$ the exchange interaction energy.  At the transition temperature, the critical scaling parameter is $1/4$, and we can extract from these data the corresponding transition temperature of 121~mK and exchange energy of -0.0066~meV, obtained by fitting the simulated curves in Fig.\ref{magT}(a). This value of $J$, which represents an effective scale, is smaller than the smallest magnitude exchange energy we calculate between neighbors, 0.016~meV, as well as the average exchange energy, -0.03~meV. The smallness of the $J$ value likely arises from the disorder of the exchange interactions.  For the standard XY model, without disorder, a fully  quantum treatment of the S~$=1/2$ system has a transition temperature of kT$_c$/$|J|$$ = 0.35$,\cite{ding1990kosterlitz} while the classical planar XY lattice has a transition at kT$_c$/$|J|$$ = 0.898$. Quantum effects depress the transition temperature and would be strongest for smaller spins. Our results were obtained using a classical Monte Carlo simulation parameterized by quantum DFT calculations, so we expect that the fully quantum result corresponding to our S~=~1 system would produce a transition temperature less than our computed 121~mK. This temperature range is compatible with a peak observed at 55~mK in a magnetic flux vs.\ temperature measurement of an 870~pH superconducting loop by Sendelbach et al.\cite{sendelbach2008magnetism}, indicating a phase transition.

 \begin{figure}[!h]
\centering
\includegraphics[width=0.96 \columnwidth]{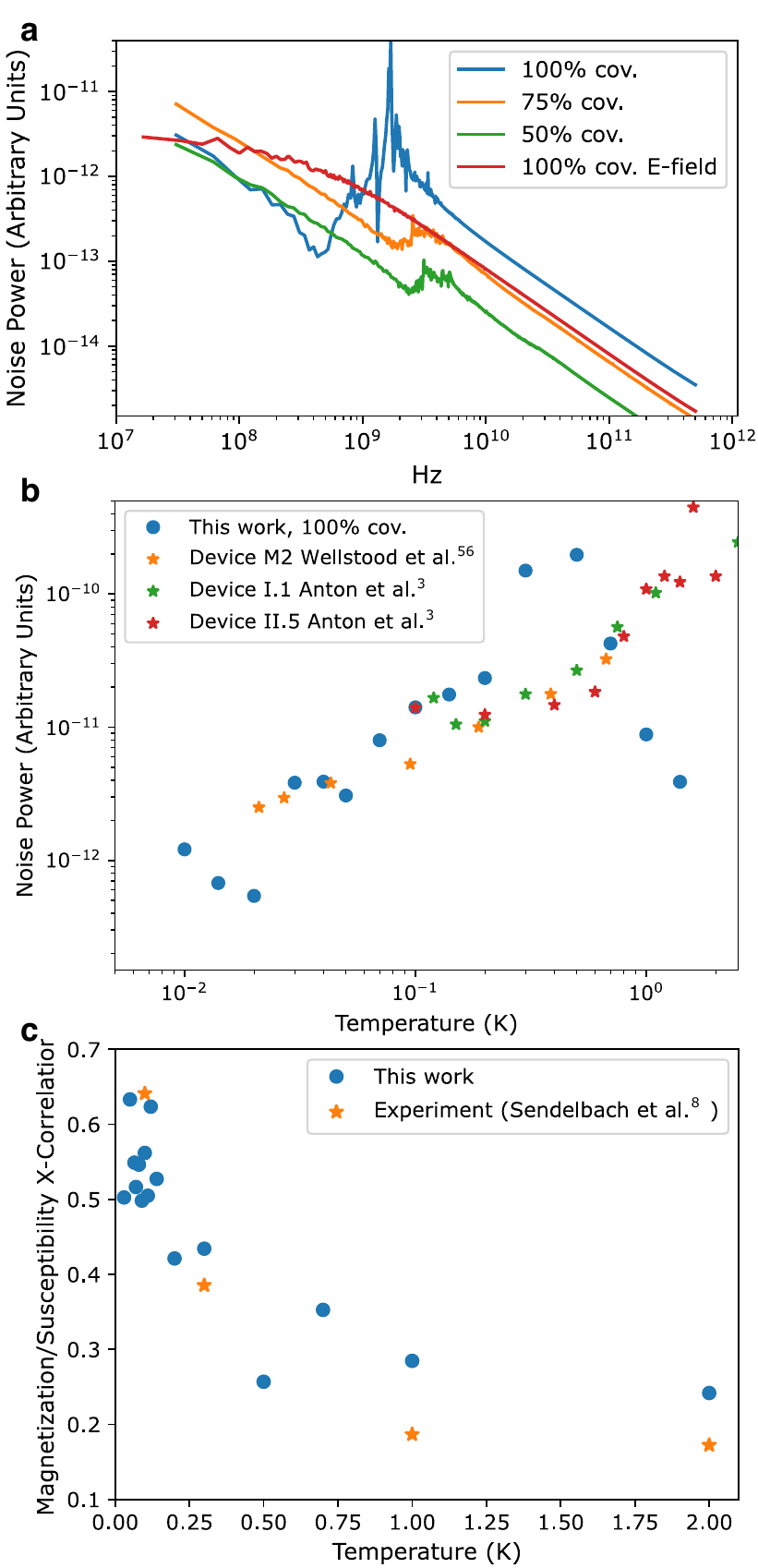}
\caption{(a) The magnetic flux noise spectrum computed for a sapphire surface covered with paramagnetic O$_2$ spins with 100\%, 75\%, and 50\% occupancy, at 0.01~K in zero applied magnetic and electric  fields, as well as with 100\% occupancy in an applied E-field. (b) Simulated magnetic flux noise at 160~MHz vs.\ temperature compared to measurements in the literature. Experimental series are each scaled to have consistent low-temperature noise magnitude, so the temperature dependency can be highlighted.
(c) Cross-correlation between magnetization and magnetic susceptibility ($S_{M\chi''} / S_M S_{\chi''} $) vs.\ temperature for our simulations averaged over frequencies between 1.5 and 2.5 MHz (blue dots) compared to measurements by Sendelbach et al. at 1 Hz\cite{sendelbach2009complex} (orange dots).
}
\label{spectrum}
\end{figure}

We next calculate the spin dynamics and flux noise spectra relevant to superconducting quantum devices using the Landau-Lifshitz-Gilbert (LLG) equation,
    \begin{equation}
  \frac{d\textbf{M}}{dt} = -\frac{\gamma}{1 + \alpha^2}\textbf{M} \times \textbf{H}_{\text{eff}} - \frac{\alpha\gamma}{1 + \alpha^2}\textbf{M} \times (\textbf{M} \times \textbf{H}_{\text{eff}}),
  \end{equation}
where $\gamma$ is the reduced gyromagnetic ratio, $\alpha$ is a dimensionless damping constant, \textbf{M} is the magnetization discretized to each pair of paramagnetic spins,  and
    \begin{equation}
\textbf{H}_{\text{eff}} = - \frac{1}{\mu_{0}}\frac{\delta H_{\text{spin}}}{\delta \textbf{M}}
  \end{equation}
  is the effective magnetic field. We perform the dynamics calculations on the ensembles of spin configurations generated with Monte Carlo simulations driven by the Hamiltonian in Eq.~\ref{Hspin} at different temperatures and values of the external magnetic field. Then, the Fourier transform of  the trajectory of the total spin yields the magnetic noise spectra, as illustrated in Fig.~\ref{spectrum}(a) for the case of zero external magnetic field and a temperature of 0.01~K. At frequencies above 6 GHz and below 30 MHz, the spectra for 50-100\% O$_2$ coverage without an applied field in Fig.~\ref{spectrum}(a) follows a $1/f^{0.8-1.0}$ scaling with frequency, as observed in numerous measurements of the magnetic flux noise in superconducting loops.\cite{slichter2012measurement}  
  There is a magnetic resonance between 1 and 2~GHz that likely originates from spin waves. This step is smaller for the 50\% and 75\% coverages compared to 100\% coverage as would be expected for spin waves due to these vacancies inhibiting collective spin motion. The effective exchange parameter fit to the spin spatial correlation described above, -0.0066 meV, corresponds to a frequency of 1.60 GHz, which is very close to the peak in the noise spectrum.

  The magnitude of the flux noise for full O$_2$ coverage at 10~MHz is calculated for a typical SQUID loop geometry (loop radius = 12 $\mu$m, line width = 0.5 $\mu$m) to be $6.6 \times10^{-9}  \phi_0/\sqrt{Hz}$ using our simulated spin fluctuation power spectral density and an analytical relation between magnetic moment and magnetic flux through the loop. This relation is reported by Anton et al.\cite{anton2013magnetic} and Bialczak et al.\cite{PhysRevLett.99.187006}  and reproduced in the Supplemental Material. For 75\% and 50\% coverages the flux noise at 10 MHz is $9.0 \times10^{-9}  \phi_0/\sqrt{Hz}$ and  $5.27 \times10^{-9}  \phi_0/\sqrt{Hz}$, respectively. These values compare well with experimental measurements from Anton et al. for the same device geometry, device I.1. After extending that measurement to 10 MHz using the frequency scaling relation fit in that work, the experimental value is $\approx 5 \times10^{-9}  \phi_0/\sqrt{Hz}$,\cite{anton2013magnetic} which in our simulations would correspond to roughly a 50\% coverage of O$_2$ molecules.

In our simulations, when the temperature of the spin system is increased from 0.01~K to 0.50~K, the flux noise at 160~MHz increases by two orders of magnitude (Fig.~\ref{spectrum}(b)). This frequency was chosen because it is at the  lower end of our simulations; typical experimental flux noise measurements are taken at lower frequencies. A similar increase in flux noise magnitude over a similar temperature range has been measured by Wellstood et al.\cite{wellstood1989hot} and Anton et al.,\cite{anton2013magnetic} as indicated in Fig.~\ref{spectrum}(b) where these measurements are reproduced and shifted in noise magnitude so that their lowest temperature noise magnitude aligns with the simulations. Interestingly, above 0.50~K, the flux noise decreases in our simulation and we note that there are measurements showing flux noise increasing and decreasing with temperature at different frequencies.\cite{anton2013magnetic} Both these simulations and measurements in the literature show a thermally driven noise process.

The cross-correlation between the magnetization and the magnetic susceptibility, $S_{M\chi''} / S_M S_{\chi''} $, has been measured by  Sendelbach et al., at 1~Hz,\cite{sendelbach2009complex} showing values as high as 0.64 at 100~mK and decreasing to 0.17 at 2~K. Non-zero values of the cross-correlation indicate a breaking of time-reversal symmetry and therefore suggest ferromagnetic clusters. Many previous simulations of the surface spin dynamics do not predict a non-zero cross-correlation,\cite{atalaya2014flux,wang2015candidate} unless imposed by explicit construction.\cite{de20191} In particular, a non-zero cross-correlation is not compatible with a spin glass, which has no net magnetization. As illustrated in Fig.~\ref{SpinLattice}(d), ferromagnetic clusters form on the paramagnetic O$_2$ surface layer. In Fig.~\ref{spectrum}(c), we plot the cross-correlation calculated from our simulations at 0.01~K to 2.0~K, zero applied field, and averaged over correlations at different frequencies between 1.5 and 2.5 MHz to obtain better statistics. These results are compared to measurements of the cross-correlation at 1~Hz,\cite{sendelbach2009complex} showing excellent agreement over the same temperature range, despite different frequencies due to limited simulation time. The higher cross-correlation numbers at low temperature indicate that the model captures the ferromagnetic spin domains, and the decrease in that cross-correlation with temperature shows that the spin interaction energy scale that emerges from this model is the same as that measured experimentally. 

\begin{figure}
\centering
\includegraphics[width=0.9 \columnwidth]{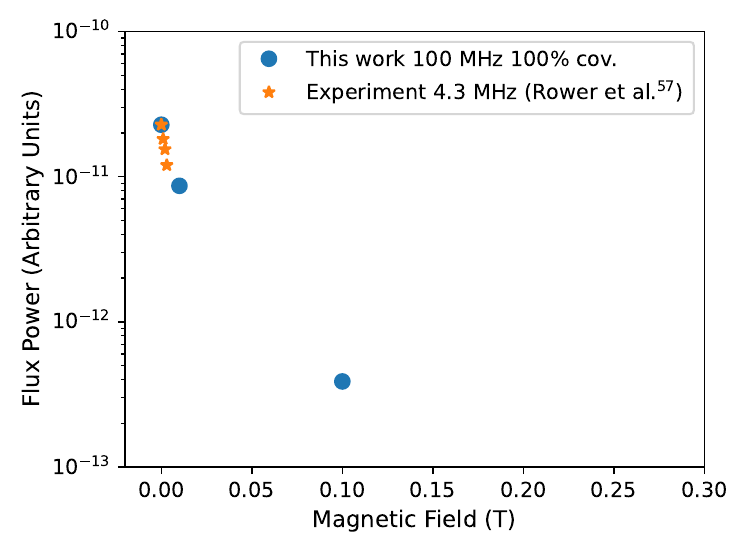}
\caption{Simulated magnetic flux noise at 100~MHz and 0.1~K vs.\ applied magnetic field. Experimental results at 4.3 MHz are included for comparison and are scaled so that the experimental zero field flux noise aligns with the simulated zero field flux noise.
}
\label{Bfield}
\end{figure}

We next consider a static external magnetic field, which at sufficient strength polarizes the spins and causes the flux noise to decrease, as shown in Fig.~\ref{Bfield}. We find that with an applied field of 0.01 (0.1)~T the magnetic flux noise at 100 MHz decreases  by a factor of 2.6 (58.7) compared to no applied field. Experimental investigations have recently found that a magnetic field of as little as 0.003~T can reduce the magnetic flux noise above 1~MHz by roughly a factor of 2.\cite{PhysRevLett.130.220602} These results show that the magnetic flux noise reducing effects of an external field are captured by our simulations at roughly the same magnitude.

These simulations track the fluctuations of electrons on paramagnetic oxygen molecules. Pairs of oxygen molecules have different electric charge distributions depending on whether they are in ferromagnetic or antiferromagnetic spin configurations. Therefore, the electric dipole moment of the oxygen molecule pair is dependent on their spin configuration and so the presence of an external electric field will cause an additional contribution to the E$_{UUUU}$-E$_{UUDD}$ energy splitting. This additional contribution is $\textbf{E} \cdot \textbf{p}$, where $\textbf{p}$ is the dipole moment difference between FM and AFM spin configurations. We calculate the electric dipole moment differences between spin up-up and spin up-down configurations for nearest neighbor oxygen molecules of different relative molecular orientations (see Fig.~\ref{SpinLattice} for a depiction of these O$_2$ orientations) and found these differences to be as large as 0.23 e\AA\ in the XY plane, with an average magnitude of 0.07 e\AA. The complete list of electric dipole moment differences from spin configuration is given in the Supplemental Material.
\begin{figure}[h!]
\centering
\includegraphics[width=0.95 \columnwidth]{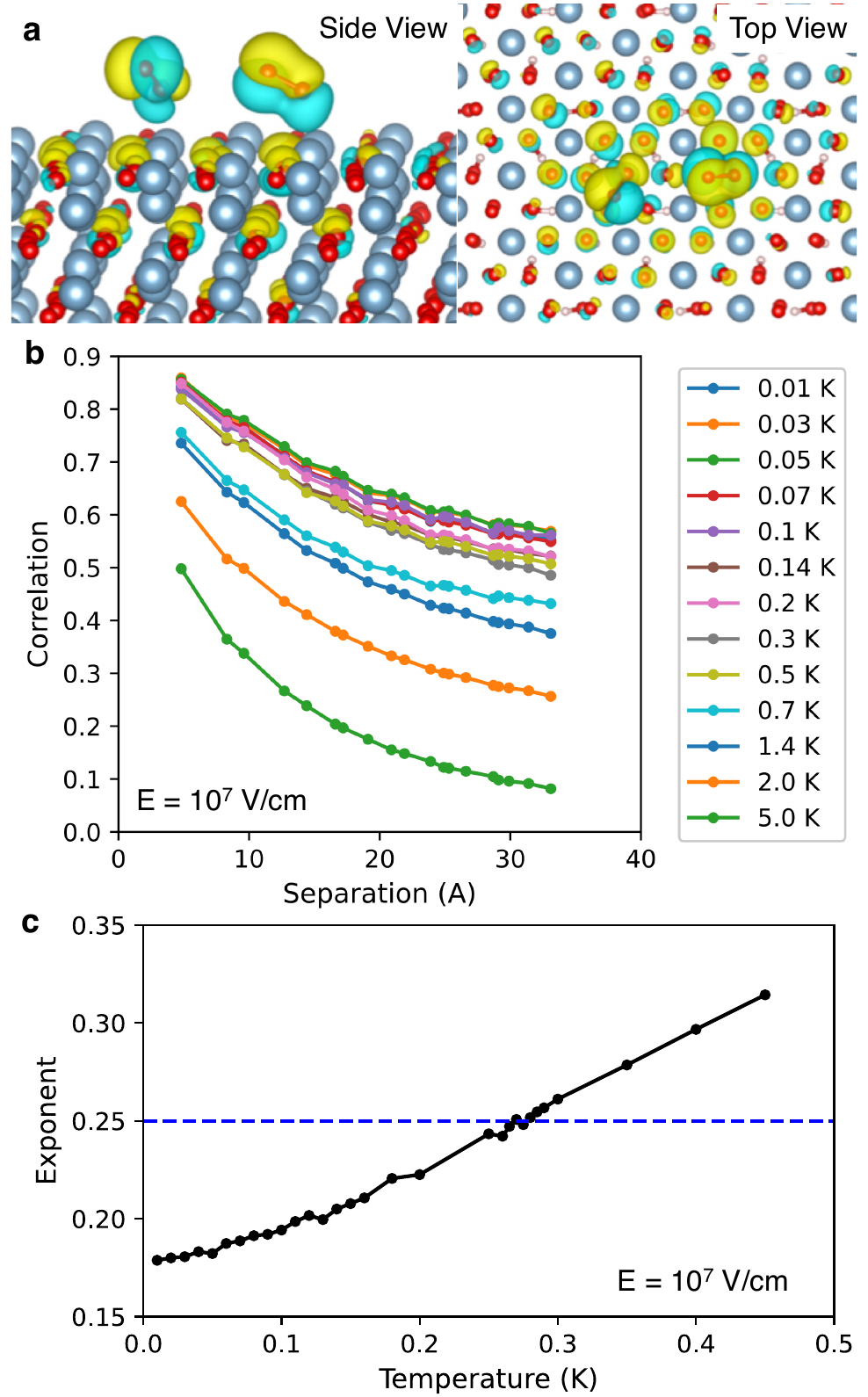}
\caption{(a) Side and top-down views of the charge density difference between antiferromagnetic (AFM) and ferromagnetic (FM) spin configurations on a particular nearest neighbor O$_2$ orientation on a (0001) sapphire surface (isosurface: 0.0025~e$/\text{\AA}^3$). Yellow (blue) denotes charge accumulation (depletion) of the AFM spin configuration compared to the FM configuration. (b) Spin-spin correlation functions for several temperatures, with an applied electric field of 10$^7$~V/cm, show the algebraic decay characteristic of BKT phases below the transition temperature. (c) For the expected algebraic scaling, $(|r - r'|/l)^{-\eta}$, $\eta = 1/4$ at the transition temperature. We plot  $\eta$ and find the transition temperature to be 276~mK.
}
\label{Emag}
\end{figure}

  Through these electric dipoles, the spin systems may couple to superconducting resonators. An example charge density difference is depicted in Fig.~\ref{Emag}(a), showing how the electron density shifts with spin state. Because of the E-field enhanced exchange interaction between nearest neighbor spins the BKT transition temperature will be altered. In Fig.~\ref{Emag}(c) we see the spin-spin correlation functions for an applied field of 10$^7$ V/cm. Compared to Fig.~\ref{magT}(a), these correlation functions are stronger to higher temperatures. As in the \textbf{E} $= 0$ case, we plot the algebraic scaling coefficient to determine the transition temperature (\ref{Emag}(d)) and find that T = 276 mK at $\eta = 1/4$, which is considerably higher than the zero-field transition temperature, 121 mK. Such a change in the spin system decreases the flux noise, as plotted in Fig.~\ref{spectrum}(a). At 10 MHz the flux noise in a typical SQUID loop geometry (loop radius = 12 $\mu$m, line width = 0.5 $\mu$m) with 100\% coverage is calculated to be $4.3 \times10^{-9}  \phi_0/\sqrt{Hz}$ when an E-field of 10$^7$ V/cm is applied, compared to $6.6 \times10^{-9}  \phi_0/\sqrt{Hz}$ with no applied field. Furthermore, the frequency scaling below 100 MHz with the applied electric field is $1/f^{0.25}$ compared to $1/f^{1.0}$ for the case without an applied electric field, so the suppression of flux noise at lower frequencies will be greater.


This model, consisting of interacting paramagnetic spins on a physically disordered lattice of O$_2$ molecules on Al$_2$O$_3$, reproduces the temperature, magnetic field, and frequency dependencies observed experimentally, as well as the overall magnitude of the flux noise and the cross-correlation of the magnetic flux and magnetic susceptibility. This strongly indicates that a surface O$_2$ layer spin system contributes to the magnetic flux noise observed in superconducting loops made from aluminum or made using a sapphire substrate. Exposure to air is common for these devices and O$_2$ molecules have relatively strong surface binding. From these simulations we learn that the arrangement of the surface O$_2$ molecules, which manifests as a network of patches each made of similarly oriented O$_2$ dimers (Fig.~\ref{SpinLattice}(b)), defines the exchange interactions between the O$_2$ unpaired spins. These exchange interactions divide the O$_2$ spins into ferromagnetic domains separated by boundaries that  host either anti-ferromagnetic interactions or a different magnitude of ferromagnetic interactions compared to the domains. It is the fluctuations of spins in these domains that generate magnetic flux noise. 

To mitigate the effects of flux noise in superconducting qubits, SQUID-based devices in general, and spin-based qubits in semiconductors or NV centers, the density of adsorbates must be reduced or the interactions between O$_2$ must be modified to reduce fluctuations. Lee et al. reported calculations showing that an NH$_2$ surface termination may yield a low magnetic flux noise surface\cite{PhysRevLett.112.017001} and Kumar et al. showed experimentally that UV and an ammonia treatment, which modifies the surface termination, each separately reduced magnetic flux noise.\cite{kumar2016origin} Future strategies to mitigate magnetic flux noise in qubits may benefit from a detailed microscopic understanding of the exchange couplings and ferromagnetic domain structure to optimally engineer the surface. We show that an applied magnetic field reduces magnetic flux noise in this system, however, the application  of this field may be limited due to the effect of magnetism on superconductivity. Alternatively, we demonstrate that a strong electric field increases exchange coupling strength, through an O$_2$ pair spin dependent electric dipole moment, and reduces magnetic flux noise. Surface modifications that change the exchange coupling in similar ways would also be predicted to reduce flux noise. The spin dependent O$_2$ electric dipole moments, along with exchange interactions that we calculate can be in the $\mu$eV range,  imply that these paramagnetic spins will couple to GHz excitations in superconducting qubits as two-level systems (TLSs) and cause loss and decoherence. Whether magnetic flux noise sources in general act as TLSs, causing dielectric loss, remains an active area of investigation important to the reduction of both magnetic flux noise and TLS dielectric loss.\cite{}

\acknowledgements
This work was performed under the auspices of the U.S. Department of Energy by Lawrence Livermore National Laboratory under Contract DE-AC52-07NA27344 and was supported by the LLNL-LDRD Program under Project No. 15-ERD-051 (prior to 2019). The QIS research at LLNL was supported by the U.S. Department of Energy, Office of Science, Basic Energy Sciences, Materials Sciences and Engineering Division (2019 onwards).

\bibliography{ref}{}
\end{document}


\title{Field-Dependent Qubit Flux Noise Simulated from Materials-Specific Disordered Exchange Interactions Between Paramagnetic Adsorbates: Supplemental Material}
\author{Keith G. Ray}
\email{ray30@llnl.gov}
\author{Yaniv Rosen}
\author{Jonathan L Dubois}  
\author{Vincenzo Lordi}  
\email{lordi2@llnl.gov}
\affiliation{Lawrence Livermore National Laboratory 7000 East Avenue, \\Livermore, CA 94550, USA}
\maketitle

\section{DFT Calculations}

To calculate the equilibrium positions, configurational energies, and exchange energies of different orientations of a pair of oxygen molecules on a (0001) Al$_2$O$_3$ surface, we use the Vienna Ab-Initio Simulation Package\cite{vasp1,vasp2,vasp3,vasp4} (VASP) with a plane-wave basis cutoff of 500 eV, pseudopotentials from the projector augmented
wave set\cite{paw2} (PAW), and the van der Waals density functional\cite{Dion:2004fk} with a ``consistently'' defined exchange functional (vdW-DF-cx).\cite{PhysRevB.89.035412} The calculations consist of pairs of O$_2$ molecules on top of an Al-terminated 3-layer thick slab of  (0001) Al$_2$O$_3$ that is passivated with hydrogen on the opposite side. The periodic slab is six surface units wide in both in-plane directions (28.70 \AA \ $\times$ 24.86 \AA) with a 14.5 \AA \ vacuum spacing (796 total atoms) and a single k-point at gamma is employed. Forces are relaxed to 0.005 eV per \AA. Each O$_2$ molecule has two unpaired spins. The spin exchange energies are calculated by starting each O$_2$ pair in both the UUUU and UUDD configuration using the MAGMOM flag and then comparing the energies when relaxed (the spin configurations persist during relaxation). Spin anisotropy is calculated by constraining the two unpaired spins on an O$_2$ molecule to directions perpendicular and parallel to the O$_2$ bond axis with the O$_2$ molecule in its equilibrium binding geometry on the (0001) Al$_2$O$_3$ surface.

\section{Monte Carlo Simulations}
\label{MC_sim}

Monte Carlo simulations for the O$_2$ orientations are carried out using the metropolis algorithm along with Eq. 1 in the main text and the energies in Table S1. Simulations start at 10$^5$ times the target temperature and then reduce the temperature smoothly for 10$^6$ steps until reaching the target temperature and then running for another 10$^6$ steps. The Monte Carlo simulations for the spin lattice proceed according to Eq. 2 in the main test and utilize spin interaction energies that are calculated with DFT for different relative O$_2$ orientations. These values are given below in Table S1. The arrangement of which spins have which interactions corresponds to the O$_2$ configuration determined with the O$_2$ configurational Monte Carlo simulation. The spin update employs a combination of the metropolis algorithm and the Wolff cluster update algorithm,\cite{PhysRevLett.62.361} where one Wolff update step is performed for every 9 metropolis steps in order to speed up equilibration. We perform 8 spin Monte Carlo simulations for each of 12 O$_2$ configurations, for a total of 96 spin MC simulations, for each choice of parameters (temperature, E-field, B-field, O$_2$ coverage) reported in the main text.

\section{Landau-Lifshitz-Gilbert Simulations}
The Landau-Lifshitz-Gilbert Simulations are carried out using the Runge--Kutta--Fehlberg Method (RKF45) with a 4 fs initial timestep, which is adaptive, and a near-zero damping factor of 0.00000001 to simulate a deterministic spin trajectory where the noise emerges from the collective spin dynamics. Simulations are each run for 100 ns and for each choice of parameters start from the spin configurations generated by the 96 spin Monte Carlo simulations described in Section \ref{MC_sim} of the Supplemental Material.

To go from the simulated spin noise power to flux noise through a superconducting loop made from a film of radius $R$, width $W$, and thickness $b$, we utiliize the relationship derived in  Bialczak et al.,\cite{PhysRevLett.99.187006}
\begin{equation}
\langle \Phi^2 \rangle \simeq \frac{2\mu^2_0}{3}\mu^2_B\sigma\frac{R}{W}\Big[ \frac{ln(2bW/\lambda^2)}{2\pi}+0.27\Big],
\end{equation}
where $\mu_0$ is the vacuum permeability, $\mu_B$ is the Bohr magneton, $\sigma$ is the density of spins, and $\lambda$ is the penetration depth. 

The imaginary part of the susceptibility is calculated using the noise power of the magnetization, $\chi'' \propto S_M$.

\section{DFT Calculated O$_2$ Configurational Energies}

The binding energy, $E_i^b(O_i)$ in Eqn. 1 in the main text, of an isolated O$_2$ on a terminating Al atom on a (0001) sapphire surface is -319.9 meV for orientations 0, 2, and 4 and -303.4 eV for orientations 1, 3, 5 (see Figure 1(a) inset in the main text). $F(O_i,O_j,\textbf{r}_i - \textbf{r}_j)$ in Eqn. 1 of the main text is given by the lower of E$_{UUUU}$ or E$_{UUDD}$ for pairs of O$_2$ molecules with a particular relative orientation, see Table S1.

\renewcommand{\arraystretch}{0.8}
\begin{table}[!h]
\begin{tabular}{c.{0.5}.{0.8}.{0.10}.{0.7}}
  \multicolumn{1}{c}{O$_2$ Relative} &  \multicolumn{1}{c}{Exchange Energy}  & \multicolumn{1}{c}{E$_{UUDD}$} & \multicolumn{1}{c}{P$_{UUUU}^X$-P$_{UUDD}^X$} & \multicolumn{1}{c}{P$_{UUUU}^Y$-P$_{UUDD}^Y$}  \\
  \multicolumn{1}{c}{Orientation} &  \multicolumn{1}{c}{E$_{UUUU}$-E$_{UUDD}$ (meV)}  & \multicolumn{1}{c}{(meV)}& \multicolumn{1}{c}{(e\AA)}& \multicolumn{1}{c}{(e\AA)}  \\
 \hline                          
$[0,0]$&-0.110&9.504&0.000163&0.004017\\
$[0,1]$&	-0.283	&24.587&0.004444	&0.004880\\
$[0,2]$&	-0.101	&10.83&-0.003987&-0.073735\\
$[0,3]$&-0.436	&23.345&-0.000587&0.000192\\
$[0,4]$&	4.063	&0.459&0.007261	&-0.035147\\
$[0,5]$&	1.249	&16.231&-0.052575&	-0.039400\\
$[1,0]$&	0.207	&20.295&0.052070&	0.120071\\
$[1,1]$&	-0.727	&34.290&0.089996&	0.142067\\
$[1,3]$&	-0.238	&30.898&0.072969&	-0.090135\\
$[1,4]$&	-2.706	&84.122&-0.197374&	0.114253\\
$[3,0]$&	-0.242	&27.607&0.002982	&-0.002043\\
$[3,1]$&	-0.324	&42.378&0.029660&	-0.104237\\
$[3,2]$&	0.098	&27.878&0.018011&	0.127203\\
$[3,3]$&	-0.286	&41.396&-0.004226	&0.012263\\
$[3,4]$&-2.066	&17.758&-0.079999	&-0.055610\\
$[3,5]$&	0.016	&40.615&-0.001385&	0.004147\\
$[4,0]$&	-0.023	&11.657&-0.007590&	-0.025681\\
$[4,1]$&-0.377	&25.893&-0.030937&	0.044864\\
$[4,3]$&-0.175	&26.205&-0.020467	&0.038991\\
$[4,4]$&	-0.154	&0.000&0.020058	&-0.012441\\  
\end{tabular}
\caption{Exchange energies for spins on nearest neighbor (NN) O$_2$ molecules (E$_{UUUU}$-E$_{UUDD}$) for the twenty symmetry distinct relative orientations of NN O$_2$ molecules. E$_{UUDD}$ of each NN O$_2$ relative orientation is also tabulated as well as the electric dipole moment differences in the $\hat{x}$ and $\hat{y}$ directions.}
\label{ExchangeE}
\end{table}

\bibliography{refsupp}{}